\documentclass[12pt]{article}
\usepackage{graphicx}
\usepackage{url}
\begin{document}
\begin{center}
     {\large\bf On the Doppler effect for photons in rotating systems}\\
    \vskip5mm
  \small  {Giuseppe Giuliani}\\
    \vskip3mm
 {Dipartimento di Fisica, Universit\`a degli Studi di Pavia, via Bassi 6, 27100 Pavia, Italy}\\
{giuseppe.giuliani@unipv.it}
\end{center}
\vskip5mm\par\noindent
{\bf Abstract}.
The analysis of the Doppler effect for photons in rotating systems, studied using the M\"ossbauer effect, confirms the general conclusions of a previous paper dedicated to experiments with photons emitted/absorbed by atoms/nuclei in inertial flight. The wave theory of light is so deeply rooted that it has been--and currently is--applied to describe phenomena in which the fundamental entities at work are discrete (photons).
The fact that the wave theory of light can describe one aspect  of these phenomena can not overshadow two issues: the corpuscular theory of light, firstly applied to the Doppler effect for photons by Schr\"odinger in 1922, is by far more complete since it describes all the features of the studied phenomena;  the wave theory can be used only when the number of photons at work is statistically significant.
The disregard of basic methodological criteria may appear as a minor fault. However, the historical development of quantum physics shows that the predominance of the wave theory of radiation, beyond its natural application domain, has hampered  the   reorientation toward the photon description of the underlying phenomena.
\vskip3mm\par\noindent
pacs {03.30.+p 76.80.+y 01.65.+g}

\section{Introduction}

In a previous paper, the Doppler effect for photons emitted/absorbed by atoms/nuclei in inertial flight has been discussed,  from Einstein's proposal of 1907 to recent, sophisticated, experiments \cite{ggdoppler}. The focus  was on the  use of the wave theory of light  and on the disregard of  the photon treatment, firstly suggested by Schr\"odinger in 1922, and based on energy and momentum conservation.
\par
It seems worth  to extend the previous analysis to the experiments with photons emitted/absorbed by nuclei without recoil (M\"ossbauer effect) on rotating devices.
The M\"ossbauer effect has been discovered in 1958. The experiments exploiting it and aimed at verifying predictions of special or general relativity began in the Sixties. M\"ossbauer-type experiments have been superseded in accuracy by experiments with unstable particles in inertial flight and with photons emitted/absorbed by atoms/nuclei in inertial flight. Nevertheless, M\"ossbauer-type experiments in accelerated systems maintain their conceptual interest. Recently, the old experiments of the Sixties have been critically revisited \cite{neweffect1} and repeated \cite{neweffect2}. These recent studies claim that the re-elaborated  experimental data obtained by Kundig \cite{kundig}, and the new ones, show that the relativistic effect in rotor experiments is accompanied by a new effect whose physical origin has to be ascertained \cite{neweffect}. Since we are interested  in the confrontation between the wave and corpuscular theory of light in the description of the relativistic effect in rotor experiments, we shall not discuss further the claim about the  new effect.
\par
This paper completes the previous one:   it may be of some interest to university teachers and researchers.
\section{Theoretical and experimental background of rotor experiments}
\subsection{Theoretical}
As in the case of photons emitted/absorbed by atoms/nuclei in inertial motion, the theoretical background for rotor experiments can be traced back to  Einstein's work.
In the paper on general relativity,
Einstein used, among others,  the following argument for illustrating the
necessity of general covariance of natural laws ($K$ is an inertial
frame, $K'$ a frame rotating with constant angular velocity with respect
to $K$):
\begin{quote}\small
 \dots
 let us imagine two
clocks of identical constitution placed, one at the origin of
coordinates, and the other at the circumference of the
circle, and both envisaged from the ``stationary'' system
$K$. By a familiar result of the special theory of relativity,
the clock at the circumference--judged from $K$--goes more
slowly than the other, because the former is in motion and
the latter at rest. An observer at the common origin of
coordinates, capable of observing the clock at the circumference
by means of light, would therefore see it lagging behind
the clock beside him. As he will not make up his mind
to let the velocity of light along the path in question depend
explicitly on the time, he will interpret his observations as
showing that the clock at the circumference ``really'' goes
more slowly than the clock at the origin. So he will be
obliged to define time in such a way that the rate of a clock
depends upon where the clock may be \cite[p. 152]{ein1916}.
 \end{quote}
This is the first appearance of the idea of comparing, at least in a `gedanken
experiment', the readings of two clocks placed on a rotating disc at different
distances from its center.
\par
Some years later, in an appendix to his volume on special and
general relativity, this idea was further developed.
Einstein considered two identical clocks on a rotating
disc, one at the center and the other at a distance $r$ from it.
The rotating reference frame is named $K'$ and the inertial (laboratory)
frame $K$:
\begin{quote}\small
A clock,  at a distance $r$ from the center of the disc, has a velocity
with respect to $K$ given by
\begin{equation}
v=\omega r
\end{equation}
where $\omega$ is the angular velocity of the disc $K'$
with respect to $K$.
If $\nu_0$ is the number of ticks of the clock per unit time
(rate of the clock) with respect to $K$ when the clock is at rest,
then the rate of the clock ($\nu$) when it moves with respect
to $K$ with velocity $v$, but at rest with respect to the disc,
will be given by, in agreement with paragraph \dots, the relation
\begin{equation}
 \nu =\nu_0 \sqrt{1-v^2/c^2}
\end{equation}
or, with a good approximation by
\begin{equation}
 \nu = \nu_0 \left( 1- {{1}\over{2}}{{v^2}\over{c^2}}\right)
\end{equation}
This expression may be written in the form
\begin{equation}
 \nu = \nu_0 \left( 1- {{1}\over{c^2}}
{{\omega^2r^2}\over{2}}\right)
\end{equation}
If we denote by $\varphi$ the effective potential difference of the centrifugal
force between the position of the clock and the center
of the disc, i.e. the work, taken as negative, that must be
 done against the centrifugal force on a unit mass for
 transporting it  from the position of the clock on the
  rotating disc
 to the center of the disc, then we shall have:
 \begin{equation}
  \varphi = - {{\omega ^2 r^2}\over{2}}
\end{equation}
 \begin{equation}
   \nu = \nu_0 \left(1+{{\varphi}\over{c^2}} \right)
\end{equation}Therefore
from this relation we see at first that  two identical clocks
 will have different rates when at different distances
 from the center of the disc.
 This result is correct also from the point of view
 of an observer rotating with the disc.
 Since, from the point of view of the disc, a gravitational
 field of potential $\varphi$ appears, the result we have
 obtained will be valid in general for gravitational
 fields.
 Furthermore, we can consider an atom emitting spectral
 lines  as a clock; therefore the following statement will
 hold:  {\em the frequency of the light absorbed or emitted
 by an atom depends on the gravitational field in which
 the atom
 find itself} \cite[p. 88-89]{ein1920}.
\end{quote}
The above Einstein's passages can reasonably be considered as the conceptual source of rotor experiments.
\subsection{Experimental}
At the end of the fifties, Rudolf M\"ossbauer discovered that a substantial fraction of  $\gamma$ photons emitted or absorbed by  nuclei in crystalline
solids  are emitted/absorbed without  energy loss due to  recoil of the emitting nucleus: the
 recoil energy is taken up by the crystal as a whole (in form of translational energy) instead of exciting the quantized  lattice vibrations.
 An elementary presentation of the effect can be found in  M\"ossbauer's Nobel Lecture \cite{mossnobel}; in \cite{recoll} M\"ossbauer recollected the paths of the discovery. The basic  formulas  can be found in the Appendix.
 \par
 M\"ossbauer's discovery was so surprising that,  as recollected by Harry Lipkin:
 \begin{quote}\small
    Rudolf M\"ossbauer's first paper was followed by an avalanche of papers,
many of which simply reproduced M\"ossbauer's experiment, obtaining exactly the
same results without adding anything new, and were published in refereed journals
as original results. This unique phenomenon reflected the general feeling in the
nuclear physics community that this experiment must be wrong and that it was
necessary to correct this error by doing the experiment right \cite[p. 4]{lipkin1}.
 \end{quote}
 The effect was unexpected in spite of the fact that the theory for explaining it had been available since many decades. As Lipkin puts it:
 \begin{quote}\small
    That photons could be scattered by atoms
in a crystal without energy loss due to recoil was basic to all work in X-ray
diffraction and crystallography. All the quantitative calculations, including the definition
and evaluation of the Debye~-~Waller factor, were well known. {\em But nobody interpreted
this as a probability that a photon could be scattered by an atom in a crystal without
energy loss due to recoil. The X-ray physicist worked entirely with the wave picture
of radiation and never thought about photons}. The Debye~-~Waller factor written as
$\exp(<-k^2x^2>)$ clearly described the loss of intensity of coherent radiation because
the atoms were not fixed at their equilibrium positions and their motion introduced
random phases into the scattered wave.
Nobody noted that scattering the X-rays involved a momentum transfer and
that coherence would be destroyed if there was any energy loss in the momentum
transfer process. They did not see that the Debye~-~Waller factor also could be
interpreted as the probability that the scattering would be elastic and not change the
quantum state of the crystal  \cite[p. 4]{lipkin1} (my italics).
 \end{quote}
 The crucial point of Lipkin's comment is that the deep rooted wave picture of radiation made it difficult to switch over to a photon picture of the phenomenon and, consequently, reinterpret the Debey~-~Waller factor in terms of  probability of a photon emission without energy loss due to recoil. Also the neglect of the photon description of the emission/absorption of radiation  by nuclei, discussed in section \ref{photons},  has contributed in hampering the reorientation toward a photon picture of the studied phenomena.
 \par
  M\"ossbauer effect has been widely used in several physics, chemistry and biophysics  fields, starting from the study  of nuclear hyperfine interactions and  the test of some predictions of special and general relativity.
\par
Around 1960, several groups began
to  test   time
dilation and gravitational red~-~shift by using M\"ossbauer effect.
 Among them, Robert Vivian Pound and
Glen A. Rebka at Harvard and John Paul Schiffer and his coworkers at
Harwell. These two teams were primarily concerned with the
gravitational red~-~shift. A historical
reconstruction of these researches can be found in \cite{hentshcel}.
Pound and Rebka realized that the emitter's and  absorber's temperatures,
if different, may overshadow the  gravitational effect \cite{rebkatemp}. They found that the measured effect--shift of the emission and absorption lines as a function of the sample's temperature--could be explained by a second order  Doppler effect due to the mean square velocity of the emitting/absorbing nuclei. Pound and Rebka did not give the calculation details; soon after, these calculations have been developed by Sherwin \cite{sherwin}.
\par
These result have been preliminary to the famous paper  on the gravitational red~-~shift \cite{pr}.   Pound and Rebka used the $14.4$ keV transition of
$^{57}Fe$. The emitter and the absorber were placed on a vertical line $22.6$ meters apart. The experimental setup was arranged in the same building used sixty years before by  Erwin Hall for studying  the free fall of bodies. In Hall's words:  ``the exact place of my experiments was the axis of the enclosed,
isolated tower of the Jefferson Physical Laboratory of Harvard
University'' \cite[p. 245]{hall}. Pound and Rebka corroborated the predicted red~-~shift with an accuracy of $10\%$. According to Pound and Rebka, the data previously obtained by Schiffer and coworkers  are unreliable owing to the neglect, among others,  of the temperature effect: ``Our experience shows that no conclusion can
be drawn from the experiment of Cranshaw et al.'' \cite[p. 340]{pr}. Five years later, Pound and Snider reached a $1\%$ accuracy \cite{ps}.
Pound and Rebka spoke of ``apparent weight of photons'' (title of the paper), thus suggesting, in some way, that they were studying the free fall of photons as, sixties years before, Hall studied the free fall of bodies.  They quoted the  paper in which Einstein  showed that if a body absorbs radiation energy this energy not only increases its inertial mass but also its gravitational one thus preserving the identities of the two \cite{ein1911}. In his argumentation, Einstein used also the assumption that an acceleration field is (locally) equivalent to a gravitational one. Pound and Snider choose a neutral title (effect of gravity on gamma radiation). They stressed their theoretical disengagement by writing:
\begin{quote}\small
    It is not our purpose here to enter into the many~-~sided
discussion of the relationship between the effect
under study and general relativity or energy conservation.
It is to be noted that no strictly relativistic concepts
are involved and the description of the effect as an ``apparent weight'' of photons is suggestive. The velocity
difference predicted is identical to that which a
material object would acquire in free fall for a time
equal to the time of flight \cite[p. 788]{ps} \footnote{If the source is at the eight $h$ from the ground and the absorber on the ground, the energy mismatch between emitter and absorber is $E_a (gh/c^2)$ where $E_a$ is the energy that the photon must have for being absorbed by the absorber. Then, in order to absorb the photon emitted by the emitter, the absorber must be moved toward the emitter with velocity $gh/c$: this is the velocity change  that a free falling body would acquire during a time interval equal to the time of flight of the photon $h/c$. In fact: $\Delta v= g \Delta t= gh/c$. }.
\end{quote}\small
And, just a few lines above:
\begin{quote}
    The present
experiment makes no direct determinations of either
frequency or wavelength. The determination of a compensating
source velocity is an exact operational description
of the experiment.
\end{quote}
In a recent paper, the interpretations of the red~-~shift experiments in static gravitational fields have been keenly discussed \cite{telegdi}.

\section{The rotor experiments of the Sixties}
Consider an emitter and an absorber of  $\gamma$ photons without recoil placed at two different distances from the center of a rigid rotating system (rotor): the $\gamma $ photons passing through the absorber
 are detected by a counter at rest in the laboratory.
 \par
 The first paper~-~by Hay, Schiffer, Cranshaw and
Egelstaff~-~appeared in
1960 \cite{schiffer}.
 The emitting and absorbing nucleus was $^{57}Fe$ and the maximum rotation speed was of  $500$ revolutions per second.
 \par
The authors write:
\begin{quote}\small
 In an adjoining paper  an experiment is described in which the change of frequency in
 a photon passing between two points of different gravitational potential has been
 measured. Einstein's principle of equivalence states that a gravitational field is
 locally indistinguishable from an accelerated system. It therefore seemed desirable
 to measure the shift in the energy of $14\,keV$ gamma rays from $^{57}Fe$ in an accelerated
 system
 \par
 [\dots]
\par
  The expected shift can be calculated in two ways. One can treat the acceleration as an
  effective gravitational field and calculate the difference in potential between
  the source and absorber, or one can obtain the same answer using the time dilatation
  of special relativity.
\end{quote}
The authors speak of  ``the change of frequency in
 a photon passing between two points of different gravitational potential'': another way of  expressing the idea of ``the apparent weight of photons'' of Pound and Rebka.
The theoretical descriptions are only referred to, without any calculation or comment.
  The  questions  not asked by this paper have been raised by Champeney
   and Moon \cite{champ}.
   They write:
\begin{quote}\small
Reporting a test of the effect of circular motion on the resonant frequency of gamma~-~ray
transition in $^{57}Fe$, Hay, Schiffer, Cranshaw and Egelstaff point out that one can
treat the acceleration as an effective gravitational field and calculate the frequency
shift from the difference of potential between source and absorber, or one can obtain
the same answer by using the time dilatation of special relativity.
\par
For their arrangement, with the source at the center and the absorber at the periphery of
the rotating system, the same result also follows from the argument that since source
and absorber have relative velocity $v\, (\ll c)$ in a direction perpendicular to the line
joining them, there exists a transverse Doppler effect giving a fractional frequency shift
$v^2/2c^2$.
\par
It is perhaps surprising that the na{\"\i}ve use of this formula, without any account been
taken of acceleration, should give the correct answer; an indication of the subtleties
that may be involved is obtained by considering source and absorber to move on the same
circle, e.g. at opposite points on the periphery. Their pseudo~-~gravitational
potentials are  equal, so are their time~-~dilatation, yet their relative velocity is
$2v$.
\par
[\dots]
\par
Since in this laboratory we were undertaking a `source at center' experiment similar to
that of Hay {\em et al.}, we decided also to make an experimental test of the
`peripherally opposite'
 arrangement \cite[p. 350]{champ}.
\end{quote}
With respect to the paper by Hay et al., the wave Doppler effect is  considered as well. This leads to an apparent puzzle  since the
 experiment with emitter and absorber at the opposite ends
of a diameter shows that there is no frequency shift, while   a transverse Doppler frequency shift should be expected.
 Champeney and Moon stress  ``the subtleties involved''
in this null result; however, they do not discuss further this ``subtle'' point.
\par
In a following paper, Champeney, Isaak and Khan reported more accurate
measurements  taken with the source at the center and the absorber at the tip
of the rotor \cite{champ2}. They
reported also on an experiment with the source at the tip and the absorber
at the center that confirmed the asymmetry of the phenomenon, due to the position interchange
 between source and absorber.
Champeney, Isaak and Khan outlined the  basic points of a  theoretical treatment based on the wave Doppler effect and,
in commenting the agreement between theory and experiment, the authors stressed that this agreement implies that the acceleration does not influence the rate of nuclear clocks. They showed also that the same result may be obtained by applying the time dilation of moving clocks. Strangely enough, they did not stress that the formula \cite[p. 584]{champ2} for the Doppler effect, written in the laboratory inertial reference system, easily explains the `subtle point' which one of the authors was talking about in \cite{champ}.
\par
The experiments by Kundig \cite{kundig} have been, according to the critical analysis  by \cite{neweffect1}, the more accurate. In Kundig's experiments, the emitter was at the rotor center and the absorber at $9.3$ cm away from it; the rotor was rotated up to $35000$ revolutions per minute (this maximum rotation speed has been, more or less, a  common value in the experiments of the Sixties); furthermore, the resonance condition was achieved by moving periodically  the emitter toward/away from the absorber.  As in the preceding papers, the theoretical descriptions referred to--without developing them--were time dilation, wave Doppler effect and general relativity (through the weak equivalence principle). However, in Kundig's case, the conceptual background  is more intricate. In fact, beside time dilation, wave Doppler effect and weak equivalence principle, Kundig refers also to the clock paradox, considered as a peculiar effect of accelerated systems. As he puts it: ``The
frequency shift in a rotating system might be described
as the transverse Doppler effect for accelerated systems,
also known as the `clock paradox' \cite[p. 2371]{kundig}''. Furthermore, in commenting the description of rotor experiment using the weak equivalence principle, he writes: ``We thus see that the transverse Doppler effect and the
time dilatation produced by gravitation appears as two
different modes of expressing the same fact, namely that
the clock which experiences acceleration is retarded
compared to the clock at rest \cite[p. 2372]{kundig}''.
\section{Review of  theoretical descriptions\label{review}}
In the experimental papers discussed above, the theoretical descriptions are only referred to, with the noticeable exception of \cite{champ2}.
In this section, we shall present `a rational reconstruction' of these theoretical descriptions and discuss them within the confrontation of the corpuscular and the wave description of radiation.
 \subsection{Doppler effect for waves\label{wave}}
 Let us start with the general formula for the relativistic Doppler effect valid for both acoustic and light signals \cite{doppleracot, doppleracot2}:
 \begin{equation}\label{dacot}
    \frac{\omega_a}{\omega_e}=\frac{1-(v_a/V)\cos (\vec V, \vec v_a)}{1-(v_e/V)\cos (\vec V, \vec v_e)}   \frac{\sqrt{1-v_e^2/c^2}}{\sqrt{1-v_a^2/c^2}}
\end{equation}
The reference system is the one in which the medium is at rest; $\vec V$ is the signal velocity; $\vec v_e$ and $\vec v_a$ the emitter and absorber velocity.
  Equation (\ref{dacot}) is obtained by  assuming that either the source emits signals of ideally null duration at a specified time interval or a periodic wave.  In the first case, the phenomenon's period  is the time interval between two consecutive signals; in the latter, it is the wave period.
In the case of light in vacuum, the reference system is an arbitrary inertial system and equation (\ref{dacot}) assumes the form:
\begin{equation}\label{dluce}
    \frac{\omega_a}{\omega_e}=\frac{1-(v_a/c)\cos (\vec c, \vec v_a)}{1-(v_e/c)\cos (\vec c, \vec v_e)}   \frac{\sqrt{1-v_e^2/c^2}}{\sqrt{1-v_a^2/c^2}}
\end{equation}
This equation is particularly useful when dealing with laboratory experiments in which both emitter and absorber are in motion. On the other end, it easily yields the standard formula for the Doppler effect in which only the relative velocity between emitter and absorber appears.
For instance,  in the reference system of the absorber $\vec v_a=0$; then equation (\ref{dluce}) reduces to:
\begin{equation}\label{dluce ord}
    \frac{\omega_a}{\omega_e}= \frac{\sqrt{1-v_e^2/c^2}}{1-(v_e/c)\cos (\vec c, \vec v_e)}
\end{equation}
Since, in this case, $\vec v_e$ is the velocity of the emitter with respect to the absorber, equation (\ref{dluce ord}) shows that the Doppler effect for light in vacuum depends only on the relative velocity between emitter and absorber.
\par
In the case of rotor experiments,  since both cosines are null in the reference frame of the laboratory, equation (\ref{dluce}) assumes the simple form:
\begin{equation}\label{dluce 2}
    \frac{\omega_a}{\omega_e}=  \frac{\sqrt{1-v_e^2/c^2}}{\sqrt{1-v_a^2/c^2}}= \frac{\sqrt{1-\Omega ^2 R_e^2/c^2}}{\sqrt{1-\Omega^2 R_a^2/c^2}}
\end{equation}
where $\Omega$ is the rotor angular velocity and $R_e, R_a$ are the distances from the rotor center of the emitter and absorber, respectively.
This equation shows immediately that the two frequencies are equal if $R_a=R_e$, i.e. if emitter and absorber are at the opposite ends of a diameter.
For small velocities ($\Omega R\ll c$) equation (\ref{dluce 2}) can be approximated by:
\begin{equation}\label{rotor onda}
    \frac{\omega_a -\omega_e}{\omega_e}\approx \frac{1}{2c^2}\Omega^2(R_a^2-R_e^2)
\end{equation}
The above description, developed by treating the gamma radiation as an electromagnetic wave explains the experimental results.
However, the M\"ossbauer effect deals with the emission/absorption of photons by nuclei without recoil. This notwithstanding, we would like to keep on with a wave description; then, we have to argue as follows. It is true that nuclei emit/absorb  photons; however, since we are dealing with a statistically significant number of photons emitted or absorbed (see below section \ref{discussion}), we can treat them as an electromagnetic wave and apply to it the standard equations for the Doppler effect in vacuum. As far as absorption in concerned, we know that this is possible only if the frequency of the emitted wave, measured in the reference frame of the emitter, is the same as that measured in the reference frame of the absorber; therefore,  we must compensate the frequency deficiency or excess by moving in a suitable way the absorber toward or away from the emitter. It works.
\subsection{Time dilation}
This description is based on the assumption that nuclei can be conceived as clocks.  If $\tau_0$ is the fundamental period of a nucleus~-~clock at rest in the laboratory, the same period as judged  by a nucleus~-~clock in flight with instantaneous velocity $v_e$ will be $\tau_0\sqrt{1-v_e^2/c^2}$. Similarly, for the nucleus~-~clock in flight with instantaneous velocity $v_a$, the period will be $\tau_0\sqrt{1-v_a^2/c^2}$. Then the numbers of strokes per unit time will be related by the formula:
\begin{equation}\label{nc}
    \nu_e=\frac{\nu_o}{\sqrt{1-v_e^2/c^2}}; \qquad \nu_a=\frac{\nu_o}{\sqrt{1-v_a^2/c^2}}
\end{equation}
Then:
\begin{equation}\label{freq}
    \frac{\nu_a}{\nu_e}=\frac{\sqrt{1-v_e^2/c^2}}{\sqrt{1-v_a^2/c^2}}=\frac{\sqrt{1-\Omega ^2 R_e^2/c^2}}{\sqrt{1-\Omega^2 R_a^2/c^2}}
\end{equation}
which coincides with equation (\ref{dluce 2}).
In spite of its  simplicity, this derivation is open to criticism. In fact, it is questionable that nuclei (or atoms) can be considered as clocks. A clock is made up of a frequency standard, a counting device and a display. In this context, we can forget the display and the counting device. But what about the frequency standard of a nucleus (or atom)? If we consider a single nucleus, it should be clear that it can not be considered as a clock: it emits or absorb photons. On the other hand, if we consider a statistically significant number of photons emitted by nuclei, then we could describe the emitted photons as a wave: in this case, the frequency standard is the  frequency of the emitted/absorbed wave related to the photon energy by the relation $\nu=E/h$.  In the case under discussion, the number of photons used in the experiments are statistically significant: then, a wave description is possible. However, the wave emitted by the nuclei of the emitter
leads us immediately to the previous case of the Doppler effect: the `time dilation explanation' coincides with that of the Doppler effect for waves; therefore, it is not a different explanation. Anyway, the `time dilation explanation' stresses the fact that transverse Doppler effect allows the measurement of the time dilation factor $\sqrt{1-v^2/c^2}$.
\subsection{The  general relativity treatment\label{general}}
The
 energy of a particle of mass $M$ at rest in a constant  gravitational field is given by \cite[p. 387]{moller}:
\begin{equation}\label{mass}
    \mathcal E= Mc^2\sqrt{1+\frac{2\phi}{c^2}}
\end{equation}
where $\phi$ is the newtonian gravitational potential.
If the particle is a nucleus in an exited state, equation (\ref{mass}) may be written as:
\begin{equation}\label{excited}
   \mathcal E=(mc^2+\Delta E) \sqrt{1+\frac{2\phi}{c^2}}
\end{equation}
where $m$ is the nucleus' mass in its fundamental state and $\Delta E$ is the energy difference between the two  levels of the nuclear transition. Then,
the energy difference between the two  levels of the nuclear transition is modified by the gravitational potential by the term $\sqrt{1+\frac{2\phi}{c^2}}$. Therefore,
the angular frequency of the nuclear transition is given by:
    \begin{equation}\label{come}
        \omega(\phi)= \omega_\infty \sqrt{1+\frac{2\phi}{c^2}}
    \end{equation}
    $\omega_\infty$ being the transition frequency without gravitational field \footnote{This treatment can also be found in \cite[p. 117]{telegdi}. A more general derivation can be found in \cite[p. 401-407]{moller}.}.
    According to the weak equivalence principle, an acceleration field is locally indistinguishable from a gravitational one. Then, in a reference frame co~-~rotating with the rotor, the energy of a photon emitted by the emitter without recoil  is be given by:
    \begin{equation}\label{es}
\varepsilon_e=\Delta E \sqrt{1-{{\Omega^2 R_e^2}\over{c^2}}}
\end{equation}
since $\phi=-1/2(\Omega^2R_e^2)$
 is the pseudo~-~gravitational potential due to  acceleration.
Analogously, the energy of the photon that can be absorbed by the absorber is given by:
\begin{equation}\label{er}
\varepsilon_a=\Delta E \sqrt{1-{{\Omega^2 R_a^2}\over{c^2}}}
\end{equation}
Therefore:
\begin{equation}
{{\varepsilon_a}\over{\varepsilon_e}} =
 {{\sqrt{1-{{\Omega^2 R_a^2}/{c^2}}}}
 \over{\sqrt{1-{{\Omega^2 R_e^2}/{c^2}}}}}
\end{equation}
In the approximation of small velocities $(\Omega R\ll c)$:
\begin{equation}\label{rsen}
{{\varepsilon_a -\varepsilon_e}\over{\varepsilon_e}}\approx
{{1}\over{2}}{{\Omega^2}\over{c^2}}(R_e^2-R_a^2)
\end{equation}
    If we compare this equation  with equation (\ref{rotor onda}), we see that they differ for a change in sign in the second member. We shall discuss this apparent contradiction in next section.
\section{A straightforward explanation in terms of photons\label{photons}}
In rotor experiments, photons are emitted by nuclei in the emitter and, in resonance conditions, they are absorbed by nuclei in the absorber. The macroscopic detector used for measuring the gamma radiation not absorbed by the absorber is a counter that counts the photons impinging on it. Therefore, we should look for a theoretical description based only on photons without any reference to waves.
As firstly shown by Schr\"odinger \cite{erwin},  the emission  of a photon by an atom or a nucleus can be appropriately described by writing down the conservation equations for energy and linear momentum. Schr\"odinger treatment can be immediately extended to the absorption case \cite[p. 1037-39]{ggdoppler} \footnote{As shown in \cite{ggdoppler}, Schr\"odinger derivation has been rapidly forgotten. During the writing of \cite{ggdoppler}, I was not aware of the paper by Dragan Red\u{z}i\'{c} on Schr\"odinger work \cite{dragan}. I thank dr. Red\u{z}i\'{c}  for having pointed out my omission. }. In the laboratory's (inertial) reference frame, the energy of the emitted photon  is given by:
\begin{equation}\label{uguale}
E_{ph} =E^0_{ph}  {{\sqrt{1-v_1^2/c^2}}\over{1 -(v_1/c)\cos \theta_1 }}
\end{equation}
with
    \begin{equation}\label{zero}
 E^0_{ph}= \Delta E \left( 1 -{{\Delta
E}\over{2E_1}}\right),\: (v_1=0)
\end{equation}
 Where: $\Delta E=  (E_1-E_2)$; $E_1$ and $E_2$ are the rest energy of the emitting particle before and after the emission  respectively;  $v_1$ the velocity of the particle before emission; $\theta_1$ the angle between $\vec v_1$ and the direction of the emitted photon; $E^0_{ph}$ is the photon energy when the emitting  particle is at rest before the  emission. Notice that: a) $\Delta E$ is the energy difference between the two quantum energy levels of the transition; b)  both $\Delta E$ and $E_{ph}^0$ are relativistic invariants since they depend only on rest energies. In the case of absorption,
 the energy $E_{ph}$ that a photon must have for being absorbed by an atom/nucleus
 is  given again by equation (\ref{uguale}), where, in this case:
    \begin{equation}\label{nu03}
E_{ph}^0 = \Delta E\left( 1 + {{\Delta E}\over{2E_1}}\right)
\end{equation}
is the energy of the photon  absorbed by an atom/nucleus at rest
before the absorption and, of course, now $\Delta E=E_2-E_1$.
\par
Then, the Doppler effect for emitted  photons is the result of the energy and momentum conservation: if the free emitting particle is at rest before emission, the energy of the emitted photon is always less than the energy difference $\Delta E$ between the two energy level of the quantum transition, owing to the fact that, during emission, the emitting particle recoils and gains kinetic energy.
On the other end, if the free emitting particle is in motion before emission, the energy of the emitted photon may be greater than $\Delta E$, the extra energy coming from a decrease of the kinetic energy of the emitting particle. In general, and more precisely, the variation of the kinetic energy of the emitting particle due to emission is given by:
\begin{equation}\label{cinetica}
    \Delta E_{kinetic}= (\gamma_2 E_2- E_2) - (\gamma_1 E_1 - E_1)
\end{equation}
Since the energy of the emitted photon is $E_{ph}= \gamma_1 E_1 - \gamma_2 E_2$ and the energy difference $\Delta E$ between the two quantum levels is  $(E_1-E_2)$, equation (\ref{cinetica}) yields:
\begin{equation}\label{cinetica2}
     E_{ph}=  \Delta E-\Delta E_{kinetic}
\end{equation}
where $\Delta E_{kinetic}\neq 0$.
Similar considerations can be developed for the energy that a photon must have for being absorbed by an atom or a nucleus. See also \cite[p. 1038-1039]{ggdoppler}.
\par
Schr\"odinger treatment can be straightly applied to rotor experiments. For zero~-~phonon emission or absorption the recoil energy of the entire crystal is given by $E_R=\Delta E^2/2Mc^2\approx 0$ where $M$ is, al least, the mass of the entire emitter/absorber \footnote{At least, since emitter and absorber is tightly bound to the rotor and the rotor to the laboratory.}. Then the energy of the emitted or absorbed photon, when the emitting/absorbing nucleus is at rest before emission/absorption, is simply $\Delta E$. Therefore, since both $\cos\theta_e$ and $\cos\theta_a$ are zero in the laboratory reference frame, the energy of the  photon  emitted by the emitter is:
\begin{equation}\label{emitter}
    \varepsilon_e= \Delta E \sqrt{1-\frac{v_e^2}{c^2}}
\end{equation}
and the energy that the photon must have for being absorbed by the absorber is:
\begin{equation}\label{absorber}
    \varepsilon_a= \Delta E \sqrt{1-\frac{v_a^2}{c^2}}
\end{equation}
If $\Omega$ is the angular velocity of the rotor and $R_e$ and $R_a$ are the distances of the emitter and the absorber from the center of the rotor, equations (\ref{emitter}) and (\ref{absorber}) become,   in the approximation for low velocities:
    \begin{eqnarray}
                                                                      \varepsilon_e &\approx&  \Delta E \left(1-\frac{\Omega^2 R_e^2}{2c^2} \right)\\
                                                                      \varepsilon_a &\approx& \Delta E \left(1-\frac{ \Omega^2 R_a^2}{2c^2}\right)
                                                                    \end{eqnarray}
 Then:
 \begin{equation}\label{rotor}
    {{\varepsilon_a -\varepsilon_e}\over{\varepsilon_e}}\approx
{{1}\over{2}}{{\Omega^2}\over{c^2}}(R_e^2-R_a^2)
 \end{equation}
 This  equation describes the experimental data. For instance, if $R_a>R_e$,  the energy that a photon must have for being absorbed by the absorber is smaller than the energy of  the photon emitted by the emitter. In this case, for restoring the resonance condition, the absorber must be moved away from the emitter, thus compensating by first order Doppler effect the energy mismatch.
 \par
 Equation (\ref{rotor}) coincides with equation (\ref{rsen}) of the general relativity treatment; both differ from equation (\ref{rotor onda}) of the wave treatment, for a change in sign of the second member.
 We shall explain this apparent contradiction by comparing equation (\ref{rotor}) with equation (\ref{rotor onda}), i.e. by comparing the corpuscular description with the wave one. A similar explanation holds for the comparison of the wave and general relativity treatment.
In equation (\ref{rotor}) the photon energies are measured in the laboratory reference frame: $\varepsilon_e$ is the energy of the photon emitted by the emitter and $\varepsilon_a$ is the energy that the photon must have for being absorbed by the absorber.
Instead, in equation (\ref{rotor onda}) the  frequencies are measured in the reference frames of the emitter ($\omega_e$) and  absorber ($\omega_a$). However, the predictions implied by the two equations are the same from the operational point of view. In fact, if
 $R_a>R_e$, the angular frequency measured by the absorber in its reference frame is greater than that measured by the emitter in its reference frame. Then, for restoring the resonance condition, the absorber must be moved away from the emitter, as required by equation (\ref{rotor}). Of course, the two descriptions  are operationally equivalent also for $R_a<R_e$.
 \par
 The equivalence of the two descriptions is only operational. The physics is different. In fact,  the photon description deals with the physics of the emission and absorption process, while the wave one relies only on the frequency transformation due to the change of the reference frame. As Einstein put it in 1905:
 \begin{quote}\small
 In spite of the complete experimental confirmation of the theory applied to diffraction, reflection, refraction, dispersion, etc., it is still conceivable  that the theory of light that operates with continuous spatial functions may lead to contradictions when it is applied to the phenomena of emission and transformation of light \cite{ein5ajp}.
 \end{quote}
 Within the wave description, the contradictions evoked by Einstein are turned away by simply ignoring the emission and absorption processes.
  \par
  The photon description, explains also the temperature effect discovered by Pound and Rebka. If,  following \cite{rebkatemp}, we assume that the velocity entering  equation (\ref{rotor}) is the mean nucleus velocity,  the average value of $\vec v_1\cos \theta_1$  is zero and there in no first order effect; only the second order effect is left. Then, by expressing the mean square velocity  in terms of the crystal temperature, we get the formula used by Pound and Rebka.
  \par
 Schr\"odinger's treatment can take into account also the dependence of the transition energy $\Delta E$ on the gravitational potential. In fact, it is sufficient to rewrite, for the emission case, equation (\ref{zero}) as follows:
  \begin{equation}\label{nu+-approx}
E^0_p \approx \Delta E\left( 1 - {{\Delta E}\over{2E_1}}\right)
 \left(1+ {{\phi}\over{c^2}} \right)
\end{equation}
where $\phi$ is the gravitational potential and the $\approx$ sign is due to the approximation for  small gravitational potential.
   An analogous correction must be made for equation (\ref{nu03}) (absorption case).
     Then, the Schr\"odinger treatment  describes also the red~-~shift experiments by Pound and Rebka. In this case, it implies that the gravitational potential modifies the transition energy $\Delta E$ (difference between the two quantum levels of the transition): there is no energy loss of a photon emitted upward or energy gain of a photon emitted downward; see also \cite[p. 117-118]{telegdi}.
     \par
  Finally, the general relativity description of rotor experiments can be reduced to Schr\"odinger's treatment, if it is assumed--as implicitly or explicitly assumed in the papers of the Sixties--that the process of emission and absorption of a photon by an atom/nucleus in an accelerated system is the same as that in an inertial one. Then, it is sufficient to replace in equation (\ref{nu+-approx}) the gravitational potential $\phi$ by the pseudo~-~gravitational potential due to  acceleration.
     \section{Discussion\label{discussion}}
In  section \ref{review} and \ref{photons} it has been shown that rotor experiments can be described by three different approaches: wave theory of light, corpuscular theory of light and general relativity. However, in section \ref{photons}, it has been argued that the general relativity treatment can be included in the Schr\"odinger's treatment of the emission/absorption of a photon by an atom/nucleus. Therefore, only two descriptions are left:    the wave and corpuscular descriptions.
  \par
  When there are two theories that describe the same experiment we must use some epistemological criteria to choose between them.
  In our case, it is sufficient to use a  criterium that can reasonably be considered as widely accepted: between two  theories choose the one that describe more features of the phenomenon under scrutiny, i.e. the theory that is more complete.
The corpuscular treatment is, by far,  the more complete one. In fact, it describes the process of emission and absorption of a photon by an atom/nucleus by focusing on the physical state of the emitting or absorbing particle: in doing this, it can take into account of any possible dependence of the physical state of the emitting/absorbing particle on external physical agents like, for instance, gravitational potential.
The wave description  ignores the emission and absorption processes  with their dynamical implications and answers only  the question: given two inertial reference systems, what is the relation between the frequencies of an electromagnetic wave measured in the two systems? Therefore, it is not surprising that its application  to an experiment in which only discrete entities (photons) are at work  obscures some fundamental features of the phenomenon.
  The reconstruction of  M\"ossbauer's discovery by Lipkin quoted above is, in this respect, paradigmatic: the deeply rooted wave theory of  radiation in the description of the elastic scattering of X-rays by crystals rendered it difficult to reorient the minds toward a probabilistic  interpretation in terms of photons of the Debey~-~Waller factor, necessary for the explanation of the M\"ossbauer effect.
     \par
     The case of the M\"ossbauer effect is not the only one in which the wave description of radiation had to give way to the corpuscular one. Twentieth Century's physics has recorded several cases that have been, at the same time, fundamental landmarks in the development of quantum physics. It is worth   recalling these steps.
   We have to begin with Planck's derivation of the black body radiation law based on two breakthroughs: the use of the statistical approach to thermodynamic (summarized by Boltzman's formula for the entropy: $S=A \,ln W$) and   the hypothesis that, given $N$ harmonic resonators of frequency $\nu$ in thermal equilibrium inside a cavity, the energy should be attributed to them in terms of the basic quantity $h\nu$, where $h$ is a `constant of Nature' (1900) \cite{planck}. In 1905, Einstein introduced the concept of light quantum as a heuristic tool that, by the way, explained the fact that in the photoelectric effect there is a frequency threshold \cite{ein5ajp}. One year later, Einstein stressed that Planck's derivation of black body radiation law presupposes the idea of light quanta and the hypothesis that the energy of an oscillator is quantized \cite{ein_06}. Furthermore, Planck's starting formula for the energy density in the cavity:
     \begin{equation}\label{black}
        u(\nu) = \frac{8\pi\nu^2}{c^3}<U(\nu)>
     \end{equation}
          (where $<U(\nu)>$ is the average energy of the oscillator of frequency $\nu$) is derived by supposing that, classically, the oscillators exchange energy with the cavity radiation continuously. Then, the quantization of the term   $<U(\nu)>$ is in contradiction with the derivation of the equation containing it: a rigorous derivation of Planck formula implies the derivation of the term ${8\pi\nu^2}/{c^3}$ independently from Maxwell's theory.
          \par
              Originally, Einstein's light quanta had only energy $E=h\nu$; some ten years later, Einstein endowed them  also with a linear momentum $h\nu/c$ \cite{ein17}. The first, successful  application of  these two properties of light quanta has been Schr\"odinger's corpuscular treatment of  the Doppler effect for photons (1922).  Compton \cite{compton} and  Debye \cite{debye}, after about ten years of painstaking and frustrating use of wave theory of radiation, explained the inelastic scattering of X- rays by switching  to the photon description. Finally, in 1924, Bose outlined the main passages for a  rigorous deduction of black body radiation law by developing the statistics of light quanta contained in a cavity at thermal equilibrium \cite{bose}.
     \par
     The cases of Planck and Compton~-~Debye
    are characterized by the fact that the switching from the wave to the corpuscular theory of radiation has been forced by the  need of explaining  experimental results since long refractory to a wave description. They are clear examples of the tendency  of  maintaining  the acquired theories as long as possible, in spite of  the stubborn resistance of experimental results at being explained by them.
    In the case of the Doppler effect for photons, the switching to the photon description has been avoided by focusing only on the frequencies measured in the reference frames of the emitter and absorber.
    Usually, physicists stay clear from   epistemological issues. Exceptions to this habit have been, in some cases, the origin of  creative insights: Einstein's special theory of relativity and the light quanta hypothesis are two examples; another example is given by Bose statistics of light quanta.
        \par
  Einstein's formula $E=h\nu$ relating the energy of a light quantum to the frequency of a monochromatic  electromagnetic wave and   de Broglie's relation $\lambda=h/p$ giving the wavelength to be associated to a particle (this relation holds also for light quanta) constitute the two pillars of the so~-~called wave~-~particle duality. We shall not deal with  the  many~-~sided  question if the wave~-~particle duality is a real property of the physical world or the product of  sedimentation of theoretical layers that, perhaps, are  waiting  for a new insight.  Instead, we shall discuss  how a
     beautiful two slits (Fresnel prism) experiment with one photon at a time \cite{single} can be  described using the wave or the photon approach.
     In this experiment, the interaction of an incoming photon with the detector produces  a localized bright spot. Only when the number of photons is statistically significant (about two thousands are enough), the interference pattern begins to appear as the `sum' of the spots. Maxwell's theory can be twisted to explain the appearance of localized spots  by assuming, {\em ad hoc}, that  the probability for a photon to hit a point of the detector is proportional to the classical intensity predicted for that point. This {\em ad hoc } assumption, while  confirming the vitality of Maxwell theory, is incompatible with  its continuous nature. On the other hand, if the number of photons is statistically significant, Maxwell theory predicts the correct result, {\em independently of the fact that the photons are used one at a time or all together}. If someone shows  us a picture of the experiment performed by \cite{single}, we could not say if that picture has been obtained by  using, say $20000$ photons one at a time or all together. So far as the wave description is concerned. Instead, if we like to describe the experiment by using  the photon concept, we can refer to the elementary treatment of the electrons interference given by Feynman in his Physics Lectures \cite[p. 37-5]{pl}. The only thing we have to do is to replace electrons with photons: as recalled above,  de Broglie's formula is valid also for photons. Then, we  shall find--as stressed by Feynman \cite[p. 37-6]{pl}--that the wave and particle descriptions have a common mathematical structure, shown in table \ref{confronto}.
     \begin{table}[h]
\centering
\begin{tabular}{|c|c|}
\hline
{\bf Maxwell}         & {\bf Quantum}\\
\hline
$\lambda=c/\nu$       & $\lambda=h/p= c/\nu=hc/E$\\
\hline
Electric field $\overrightarrow{E}$ & Probability Amplitude $\Psi=Ce^{\imath\varphi}$\\
\hline
$\overrightarrow{E}=\overrightarrow{E_1}+\overrightarrow{E_2}$ & $\Psi=\Psi_1+\Psi_2$\\
\hline
Energy density $ \propto E^2$ & Probability $\propto |\Psi|^2$\\
\hline
\end{tabular}
\caption{Wave and quantum description of light interference.}
\label{confronto}
\end{table}
\par\noindent
This common mathematical structure explains why the wave description predicts the correct result when the number of photons is statistically significant. However, this common mathematical structure does not tell us why a statistically significant number of photons can be treated as a wave, independently of the fact that they are used one at a time or all together. This unanswered question can be put in another way. All our experimental and theoretical acquired knowledge tells us that light is composed by ``a finite number of energy quanta which are localized at points in space,
which move without dividing, and which can only be produced and absorbed
as complete units'' as Einstein boldly predicted in 1905 \cite[p. 368]{ein5ajp}.  Then, why in the circumstances of many enough photons a wave description is possible? We do not know the answer: we only knows that it works.
\par
One further comment.
The wave description of the two slits experiment can not say which is the path of the energy between the two slits and the detector,  just as in the quantum description we can not say through which slit the photon has passed: this is another aspect of the common mathematical structure of the two descriptions.
 \section{Conclusions}
The wave theory of light is so deep rooted that it has been--and currently is--applied to describe phenomena in which the fundamental entities at work are discrete (photons):
the Doppler effect for photons, studied in \cite{ggdoppler} and in the present paper, is, from this viewpoint, paradigmatic.
The fact that the wave theory of light can describe one aspect  of these phenomena can not overshadow two issues: the corpuscular theory of light, firstly applied to the Doppler effect for photons by Schr\"odinger in 1922, is by far more complete since it describes all the features of the studied phenomena;  the wave theory can be used only when the number of photons at work is statistically significant.  The disregard of basic methodological criteria may appear as a minor fault. However, the historical development of quantum physics shows that the predominance of the wave theory of radiation, beyond its natural application domain, has hampered  the   reorientation toward the photon description of the underlying phenomena.
\section{Appendix: basic formulas of the M\"ossbauer effect}
\small
In the case of emission of a photon by a nucleus in a crystal, the recoil energy $E_R$ delivered to the crystal can be written as:
 \begin{equation}\label{energy}
    E_{R}=E_{transl}+ E_{vib}
 \end{equation}
 The emission probability  of  a photon without excitation of lattice vibrations (zero phonon process, in the language of solid state physics) is given by the (reinterpreted)  Debey-Waller factor $f$:
 \begin{equation}\label{DW}
    f= \exp \left[  - \frac{E_R}{k_B \Theta_D}\left(\frac{3}{2}+\frac{\pi^2 T^2}{\Theta_D^2}  \right) \right]
 \end{equation}
 if we assume Debye model for lattice vibrations and $T\ll\Theta_D $. $\Theta_D$ is the Debye temperature defined as $k_B\Theta_D= \hbar \omega_D$, $\omega_D$ being the maximum angular frequency of Debye model.
 From (\ref{DW}) we see that the emission probability  of  a photon without excitation of lattice vibrations increases with decreasing recoil energy and temperature and with increasing Debye temperature. As $T\rightarrow 0$, $f$ becomes constant and assumes the value:
 \begin{equation}\label{DW0}
    f\approx  \exp\left( - \frac{3 E_R}{2k_B\Theta_D} \right)
 \end{equation}
 The zero phonon  emission line has a natural line-width (width at half height) given by $\Gamma_0=\hbar/\tau_n$ where $\tau_n$ is the lifetime of the excited state of the nuclear transition. For the $14.4$ keV transition of $^{57}Fe$, $\Gamma_0=4.7\times 10^{-9}$ eV and the energy resolution is $\Gamma_0/E_\gamma=3.26\times 10^{-13}$. Now suppose that the zero phonon line emitted by the emitter is absorbed by the absorber. If we move the emitter toward/away  the absorber, owing to first order Doppler effect, the absorber will be driven off resonance when the two zero-phonon lines of emitter and absorber no longer overlap. Then, by plotting the measured absorption as a function of the emitter velocity, the absorption curve will have a half width of about $2\Gamma_0$.  The velocity $v=c\Gamma_0/E_\gamma$ is the one necessary to go from the maximum of the measured absorption line to its half value. For the $14.4$ keV transition of
$^{57}Fe$, $v= 9.8 \times 10^{-2} \,\rm{mm\, s^{-1}}$.
\par
M\"ossabuer obtained his first result with the $129$ keV line of $^{191}Ir$ whose properties are somewhat worse than those of  $14.4$ keV transition of $^{57}Fe$: larger natural line width  ($4.6\times 10^{-6}$ eV), lower energy resolution ($3.5 \times 10^{-11}$) and lower fraction of  emitted photons without recoil owing to their larger energy (see equation (\ref{DW})). The better performance of the $14.4$ keV transition of $^{57}Fe$ is the reason why successive experiments have been performed using this transition.
\par
  If, for some reason, the energy of the photons emitted through a zero phonon transition is altered by an amount larger than the natural half width, these photons can not be absorbed by the absorber. In order to restore  resonance the absorber, for instance, must be moved toward/away from the emitter with an appropriate velocity, thus compensating through first order Doppler effect the energy mismatch.

 \end{document}